\documentclass[aps,prb,reprint,groupedaddress]{revtex4-2}

\usepackage{float}
\usepackage{amsmath,amssymb,amsfonts}
\usepackage{bm}
\usepackage{graphicx}
\usepackage{epstopdf}
\usepackage{color}
\usepackage{hyperref}
\usepackage{wasysym}
\usepackage{multirow}
\usepackage{mathtools}
\usepackage{stmaryrd}
\usepackage{algorithm}
\usepackage{algpseudocode}

\hypersetup{colorlinks,citecolor=blue,linkcolor=blue,urlcolor=blue}

\definecolor{myred}{RGB}{170,0,0}
\newcommand{\ket}[1]{\ensuremath{|#1\rangle}}
\newcommand{\bra}[1]{\ensuremath{\langle#1|}}
\newcommand{\rket}[1]{\ensuremath{\left|#1\right\rangle}}

\DeclarePairedDelimiterX{\norm}[1]{\lVert}{\rVert}{#1}

\begin{document}

\title{Motif magnetism and quantum many-body scars}

\author{Eli Chertkov and Bryan K. Clark}
\affiliation{Institute for Condensed Matter Theory and IQUIST and Department of Physics, University of Illinois at Urbana-Champaign, Urbana, Illinois 61801, USA}

\begin{abstract}
We generally expect quantum systems to thermalize and satisfy the eigenstate thermalization hypothesis (ETH), which states that finite energy density eigenstates are thermal.
However, some systems, such as many-body localized systems and systems with quantum many-body scars, violate ETH and have high-energy athermal eigenstates.
In systems with scars, most eigenstates thermalize, but a few atypical scar states do not. Scar states can give rise to a periodic revival when time-evolving particular initial product states, which can be detected experimentally.
Recently, a family of spin Hamiltonians was found with magnetically ordered 3-colored eigenstates that are quantum many-body scars  [Lee et al. Phys.  Rev.  B 101, 241111(2020)]. These models can be realized in any lattice that can be tiled by triangles, such as the triangular or kagome lattices, and have been shown to have close connections to the physics of quantum spin liquids in the Heisenberg kagome antiferromagnet.
In this work, we introduce a generalized family of $n$-colored Hamiltonians with ``spiral colored'' eigenstates made from $n$-spin motifs such as polygons or polyhedra. We show how these models can be realized in many different lattice geometries and provide numerical evidence that they can exhibit quantum many-body scars with periodic revivals that can be observed by time-evolving simple product states.
The simple structure of these Hamiltonians makes them promising candidates for future experimental studies of quantum many-body scars.
\end{abstract}

\maketitle

\section{Introduction}

A subsystem of an isolated quantum system tends to equilibrate with the rest of the system, a process known as thermalization. Thermalization generally occurs in quantum systems obeying the eigenstate thermalization hypothesis (ETH) \cite{Deutsch1991,Srednicki1994,Rigol2008,DAlessio2016,Deutsch2018}, which states that energy eigenstates with finite energy density appear thermal and so exhibit thermal properties such as volume-law entanglement entropy. While ETH holds for many quantum systems, known as ergodic systems, it can sometimes be violated. For example, in many-body localized (MBL) systems \cite{Basko2006,Oganesyan2007,Pal2010,Nandkishore2015,Abanin2017,Abanin2019}, systems with strong disorder and interactions, all or a significant fraction of eigenstates appear athermal, e.g., exhibit area-law entanglement entropy. In certain non-disordered systems with local constraints, such as the one-dimensional chain of Rydberg atoms recently studied in the experiment in Ref.~\onlinecite{Bernien2017}, most eigenstates are thermal and yet a set of measure zero eigenstates in the middle of the spectrum, known as quantum many-body scars, exhibit athermal behavior.

Following the experimental observation in Ref.~\onlinecite{Bernien2017}, there has been an extensive search for systems with quantum many-body scars\cite{Bernien2017,Naoto2017,Turner2018a,Turner2018b,Moudgalya2018a,Moudgalya2018b,Ho2019,Lin2019scars,Khemani2019,Choi2019,Bull2019,Schecter2019,Ok2019,Iadecola2019,Iadecola2020,Sala2020,Khemani2020,Lin2020a_scars,Lee2020,Moudgalya2020a,Michailidis2020,Yu2018,Moudgalyay2020b,Lin2020b_scars,McClarty2020,Wildeboer2020,MondragonShem2020}. Some scar states have a high overlap with particular product states and have constant energy spacings between them. These properties imply that the time-evolution of these product states exhibits periodic revivals, a clearly athermal behavior which can be observed experimentally. In contrast, the time-evolution of a generic product state will lead to thermalization, since it has vanishing overlap with scar states. It is useful to find examples of simple Hamiltonians with such scar states so that they can be realized in experiments.

In a series of recent works \cite{Changlani2017,Changlani2019,Lee2020,Pal2021}, a large family of spin Hamiltonians with exactly known ``projected 3-colored'' eigenstates has been found with a number of interesting properties, such as large eigenstate degeneracy and quantum many-body scars \cite{Lee2020}. This family of Hamiltonians, which have fine-tuned $XXZ$ interactions, is built out of a triangle motif. By tiling together triangles, these Hamiltonian can be built on geometries such as the triangular, kagome, hyperkagome, and pyrochlore lattices \cite{Changlani2017}. The 3-colored Hamiltonian on the kagome lattice appears to be at the intersection of many different phases, which include ordered phases and potential quantum spin liquid phases \cite{Changlani2017}, suggesting that this family of Hamiltonians includes special points in phase diagrams that connect different phases of matter. When the triangle motifs are added together with positive coefficients, the resulting Hamiltonians are frustration-free and the projected 3-colored states are exact ground states, but when the triangle motifs are added together with positive and negative coefficients, the resulting Hamiltonians are frustrated and the projected 3-colored states are exact eigenstates in the middle of the spectrum. These states have logarithmic in volume entanglement entropy, large overlap with certain product states, and constant energy spacings between them, making them quantum many-body scars that exhibit periodic revivals \cite{Lee2020}.

In this work, we generalize the 3-colored Hamiltonians into a larger class of $n$-colored Hamiltonians. We show that the $n$-colored Hamiltonians have degenerate ``projected spiral colored'' eigenstates, which can be made into ground states or quantum many-body scars that exhibit periodic revivals. These Hamiltonians, which involve nearest and next-nearest neighbor $XXZ$ interactions \cite{Batista2009,Changlani2017}, are built out of $n$-spin motifs, which can be shaped as polygons or polyhedra. We give examples of many types of $n$-colored lattices that can be built out of such motifs, such as Archmidean tilings and quasicrystal lattices. We provide numerical evidence for the existence of quantum many-body scars with simple periodic revivals in these systems.

\section{Motif Construction}

In this work, we construct Hamiltonians built out of motifs. A motif is a spatial pattern of spins that will be repeated to construct larger structures. In our case, we consider motifs that are polygons and polyhedra with spins at the vertices. For each polygon or polyhedron motif, we associate a motif Hamiltonian. First, we explain how we construct the motif Hamiltonians from a previously studied spin chain Hamiltonian. Then, we describe the nature of the ground states in these motif Hamiltonians, which exhibit ``spiral colored'' magnetic order.

\subsection{Motif Hamiltonians}

Consider a small, periodic chain of $L_p$ spins. When the spins in the chain are evenly arranged on a circle, we see that they form a regular polygon. For example, a periodic chain of $L_p=6$ spins forms a regular hexagon. Therefore, a finite length spin chain defines a polygon motif and a spin chain Hamiltonian can be interpreted as a motif Hamiltonian on a polygon motif. Similarly, polyhedron motif Hamiltonians can also be built out of spin chain Hamiltonians by arranging the spins appropriately.

The spin chain Hamiltonians that we use to define our motif Hamiltonians were originally studied by Refs.~\onlinecite{Batista2009,Batista2012}. These Hamiltonians take the form
\begin{align}
\hat{H}_p[Q_p] &= \sum_{r=1,2}\sum_{i=1}^{L_p} J^{XY}_{r}(\hat{\sigma}_i^x \hat{\sigma}_{i+r}^x + \hat{\sigma}_i^y \hat{\sigma}_{i+r}^y) + J^Z_{r} \hat{\sigma}_i^z \hat{\sigma}_{i+r}^z \label{eq:HQ}
\end{align}
where $\hat{\sigma}^x_i,\hat{\sigma}^y_i,\hat{\sigma}^z_i$ are Pauli matrices on site $i$, the boundary conditions are periodic, $L_p$ is the length of the chain, $Q_p=2\pi m/L_p$ for $m=0,\ldots,L_p-1$, and the coupling constants satisfy
\begin{align}
\frac{J^{Z}_1}{J^{XY}_1} &= \cos(Q_p) = -\frac{J^{XY}_1}{4J^{XY}_2} \nonumber \\
\frac{J^{Z}_2}{J^{XY}_2} &= \cos(2Q_p) = -1+\frac{1}{8}\left(\frac{J^{XY}_1}{J^{XY}_2}\right)^2 \label{eq:Qconstraints}
\end{align}
and $J^{XY}_2 > 0$. Note that these Hamiltonians have nearest and next-nearest neighbor exchange and Ising interactions of a particular form. For later convenience, following Ref.~\onlinecite{Changlani2017}, we will refer to a two-site 
\begin{align}
J^{XY}(\hat{\sigma}^x_i \hat{\sigma}^x_j + \hat{\sigma}^y_i \hat{\sigma}^y_j) + J^Z \hat{\sigma}^z_i \hat{\sigma}^z_j \label{eq:Hxxz_bond}
\end{align}
interaction with coupling constants $J^Z/J^{XY}=\alpha$ and $J^{XY}>0$ ($J^{XY}<0$) as an antiferromagnetic (ferromagnetic) $XXZ[\alpha]$ bond.

We call the Hamiltonians of Eq.~(\ref{eq:HQ}) with fixed finite $L_p$ \emph{motif Hamiltonians} because of their role in our construction of Hamiltonians built out of motifs. For each polygon or polyhedron motif $p$, we will associate a potentially different motif Hamiltonian $\hat{H}_p[Q_p]$.

Note that the Hamiltonians in Eq.~(\ref{eq:HQ}) commute with the global $z$-magnetization operator $\hat{S}_z=\sum_i \hat{\sigma}^z_i$.

\subsection{Motif Hamiltonian ground states: spiral colored states and projected spiral colored states}

The Hamiltonians in Eq.~(\ref{eq:HQ}) have many degenerate ground states\cite{Batista2009}. In this degenerate ground state space, there are product states of the form
\begin{align}
\ket{\psi_p[Q_p,\phi_p]} &\equiv \bigotimes_{j=0}^{Lp-1}\frac{1}{\sqrt{2}}\left(\rket{\uparrow}_{j+1} + e^{i(Q_p j+\phi_p)}\rket{\downarrow}_{j+1}\right) \nonumber \\
&= \bigotimes_{j=1}^{L_p} \ket{\gamma_j}_{j} \label{eq:psiQ}
\end{align}
where $\phi_p$ is an arbitrary angle \cite{Batista2009}. These states, which Ref.~\onlinecite{Batista2009} referred to as ``canted spiral'' states, are spin waves with the spin at site $j$ pointing in the $xy$-plane of the Bloch sphere at an angle of $Q_p j+\phi_p$ with the $x$-axis. Note that the phases $Q_p j=2\pi m j/L_p$ are $N_c$-th primitive roots of unity, where $N_c=L_p/\gcd(m,L_p)$. Therefore, each spin state can be labeled with a ``color'' $\gamma_j = mjN_c/L_p \textrm{ mod } N_c \in [0, N_c-1]$ that corresponds to one of the $N_c$ roots of unity. For example, for $L_p=3, Q_p=2\pi/3, \phi_p=0$, the local $\ket{\gamma}$ states are
\begin{gather*}
\ket{0} = (\rket{\uparrow} + \rket{\downarrow})/\sqrt{2}, \quad \ket{1} = (\rket{\uparrow} + \omega \rket{\downarrow})/\sqrt{2},  \\
\quad \ket{2} = (\rket{\uparrow} + \omega^2 \rket{\downarrow})/\sqrt{2},
\end{gather*}
where $\omega = e^{i2\pi/3}$, and have been called ``red'', ``green'', and ``blue'' states in previous studies\cite{Changlani2017,Changlani2019,Lee2020,Pal2021}. Therefore, we refer to Eq.~(\ref{eq:psiQ}) as ``spiral colored'' states. These states are ground states of Eq.~(\ref{eq:HQ}) with energy 
\begin{align}
E_p \equiv L_p (J_1^{Z} + J_2^{Z})=-L_p (2 + \cos(2Q_p)) J_2^{XY}. \label{eq:Ep}
\end{align}

Due to the global $z$-magnetization symmetry of the Hamiltonian, we can construct other states in the ground state space by taking projections of the product state in Eq.~(\ref{eq:psiQ}). In particular, the projected spiral colored states
\begin{align}
\hat{P}_{S_z}\ket{\psi_p[Q_p,\phi_p]},
\end{align}
where $\hat{P}_{S_z}$ projects onto the $S_z$ quantum number sector of $\hat{S}_z$, are also exact ground states of the motif Hamiltonian in Eq.~(\ref{eq:HQ}). Other degenerate ground states of Eq.~(\ref{eq:HQ}) with anyonic properties are described in Ref.~\onlinecite{Batista2012}.

\begin{figure}
\begin{center}
\includegraphics[width=0.45\textwidth]{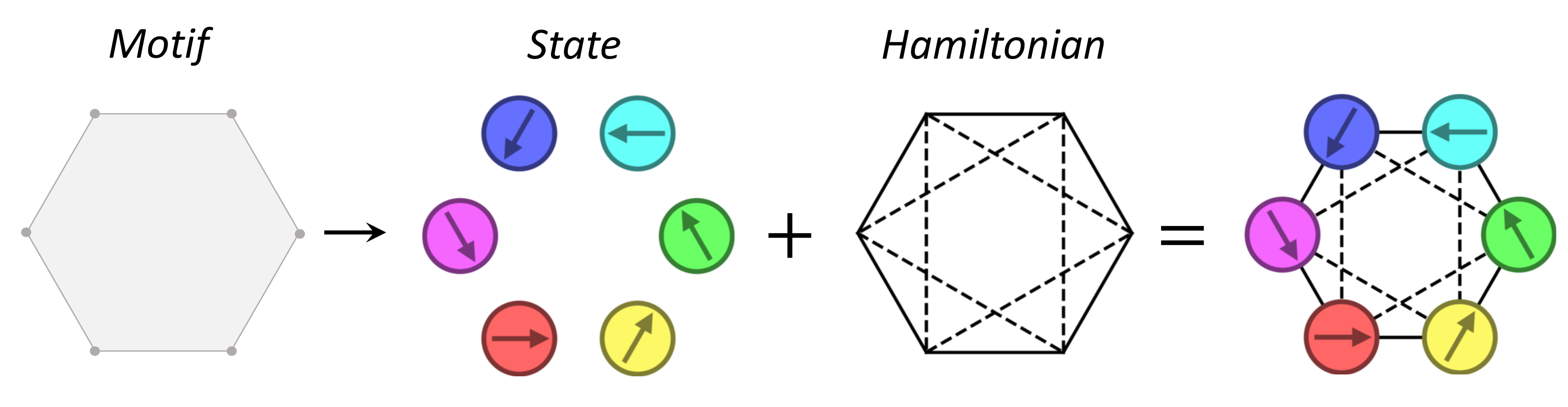}
\end{center}
\caption{An example of a motif state (Eq.~(\ref{eq:psiQ})) and motif Hamiltonian (Eq.~(\ref{eq:HQ})) for a hexagon motif. The solid (dashed) lines indicate nearest (next-nearest) neighbor spin-spin interactions in the motif Hamiltonian. The colored arrows indicate the orientation of the spins in the spiral colored states on the motifs.} \label{fig:hexagon_motif}
\end{figure}

\begin{figure}
\begin{center}
\includegraphics[width=0.45\textwidth]{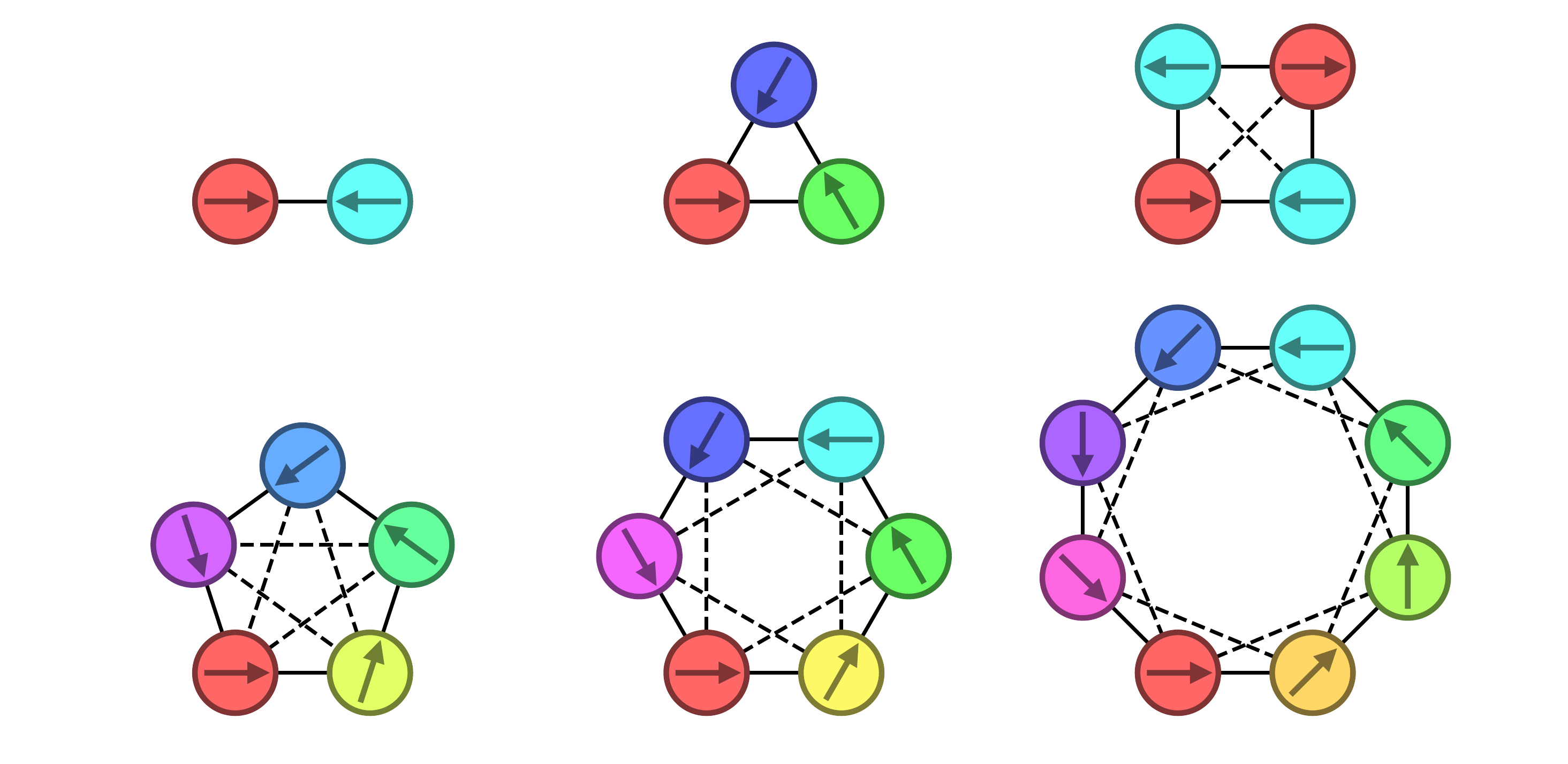}
\end{center}
\caption{Examples of motif states and motif Hamiltonians for motifs of $L_p=2,3,4,5,6,8$ spin regular polygons. The motif states $\ket{\psi_p[Q_p,\phi_p]}$ have $Q_p=2\pi/L_p,\phi_p=0$ and so are spiral colored states with $N_c=L_p$ colors, except for the square motif which has $Q_p=\pi$ and is 2-colored.} \label{fig:polygons}
\end{figure}

For our purposes, we will only consider motifs with a small number of spins $L_p$. Fig.~\ref{fig:hexagon_motif} shows an example of an $L_p=6$ motif shaped as a hexagon, depicting both the (unprojected) spiral colored state of Eq.~(\ref{eq:psiQ}) and the motif Hamiltonian of Eq.~(\ref{eq:HQ}). Fig.~\ref{fig:polygons} shows $L_p=2,3,4,5,6,8$ spin motifs shaped as regular polygons. Table~\ref{tab:HQ} lists the coupling constants of the motif Hamiltonians\footnote{Note that for $L_p=2,3$ the next-nearest neighbor bonds are actually nearest neighbor bonds and for $L_p=4$ the next-nearest neighbor bonds are doubled.} and the number of colors for different $L_p$ and $Q_p$. We point out that the Hamiltonians with 3-colored ground states made from triangular motifs, referred to as the $XXZ0=XXZ[-1/2]$ model in previous works \cite{Essafi2016,Changlani2017,Chertkov2018,Changlani2019,Lee2020,Pal2021}, is a special case ($L_p=3,Q_p=\pm 2\pi/3$) of the more general family of polygonal motif Hamiltonians with spiral colored ground states considered in this work.

\begin{table*}
\centering
\resizebox{2\columnwidth}{!}{%
\begin{tabular}{|l||c|c||c|c|c|c|} 
\hline
& $Q_p$ & $N_c$ & $J^{XY}_1$ & $J^{Z}_1$ & $J^{XY}_2$ & $J^{Z}_2$ \\ 
\hline \hline
\multirow{2}{*}{$L_p=2$} & $0$ & $1$ & $-1$ & $-1$ & --- & --- \\ 
& $\pi$ & $2$ & $1$ & $-1$ & --- & ---  \\ 

\hline \hline
\multirow{3}{*}{$L_p=3$} & $0$ & $1$ & $-1$ & $-1$ & --- & --- \\ 
& $\pm 2\pi/3$ & $3$ & $1$ & $-1/2$ & --- & --- \\ 

\hline \hline
\multirow{4}{*}{$L_p=4$} & $0$ & $1$ & $-1$ & $-1$ & $1/2$ & $1/2$ \\ 
& $\pm \pi/2$ & $4$ & $0$ & $0$ & $1$ & $-1$ \\ 
& $\pi$ & $2$ & $1$ & $-1$ & $1/2$ & $1/2$ \\ 

\hline\hline
\multirow{5}{*}{$L_p=5$} & $0$ & $1$ & $-1$ & $-1$ & $1/4$ & $1/4$ \\ 
	& $\pm 2\pi/5$ & $5$ & $-1$ & $-1/2\varphi$ & $\varphi/2$ & $-\varphi^2/4$ \\ 
& $\pm 4\pi/5$ & $5$ & $1$ & $-\varphi/2$ & $1/2\varphi$ & $1/4\varphi^2$ \\ 

\hline\hline
\multirow{3}{*}{$L_p=6$} & $0$ & $1$ & $-1$ & $-1$ & $1/4$ & $1/4$ \\ 
& $\pm \pi/3$ & $6$ & $-1$ & $-1/2$ & $1/2$ & $-1/4$ \\ 
& $\pm 2\pi/3$ & $3$ & $1$ & $-1/2$ & $1/2$ & $-1/4$ \\ 
& $\pi$ & $2$ & $1$ & $-1$ & $1/4$ & $1/4$ \\ 

\hline\hline
\multirow{4}{*}{$L_p=8$} & $0$ & $1$ & $-1$ & $-1$ & $1/4$ & $1/4$ \\ 
& $\pm \pi/4$ & $8$ & $-1$ & $-1/\sqrt{2}$ & $1/\sqrt{8}$ & 0 \\ 
& $\pm \pi/2$ & $4$ & $0$ & $0$ & $1$ & $-1$ \\ 
& $\pm 3\pi/4$ & $8$ & $1$ & $-1/\sqrt{2}$ & $1/\sqrt{8}$ & $0$ \\ 
& $\pi$ & $2$ & $1$ & $-1$ & $1/4$ & $1/4$ \\ 

\hline\hline
$L_p\geq 5$ & $2\pi m/L_p$ & $L_p/\gcd(m,L_p)$ & $-\textrm{sign}(\cos(Q_p))$ & $-|\cos(Q_p)|$ & $1/(4|\cos(Q_p)|)$ & $-|\cos(Q_p)|(\tan(Q_p)^2-1)/4$ \\ 
\hline
\end{tabular}
}	
\caption{The coupling constants $J_1^{XY},J_1^{Z},J_2^{XY},J_2^{Z}$ of motif Hamiltonians (Eq.~(\ref{eq:HQ})) that have spiral colored state ground states (Eq.~(\ref{eq:psiQ})), scaled so that $|J^{XY}_1|=1$ and $J^{XY}_2 > 0$. The spiral colored state wave vector is $Q_p=\pm 2\pi m/L_p$ where $m=0,1,\ldots,\lfloor L_p/2\rfloor$ and $N_c$ is the number of distinct colors in the spiral colored state. The constant $\varphi=(1+\sqrt{5})/2$ is the golden ratio.} \label{tab:HQ}
\end{table*}

Finally, it is important to note that $\ket{\psi_p[Q_p,\phi_p]}$ and $\ket{\psi_p[-Q_p,\phi_p']}$ for generic\footnote{If $Q_p=0$ or $\pi$ and $\phi_p=\phi_p'$, then $\ket{\psi_p[Q_p,\phi_p]}\propto\ket{\psi_p[-Q_p,\phi_p']}$ are proportional so they are not linearly independent.} $Q_p,\phi_p,$ and $\phi_p'$ are linearly independent, but non-orthogonal, ground states of $\hat{H}_p$.

\begin{figure}
\begin{center}
\includegraphics[width=0.45\textwidth]{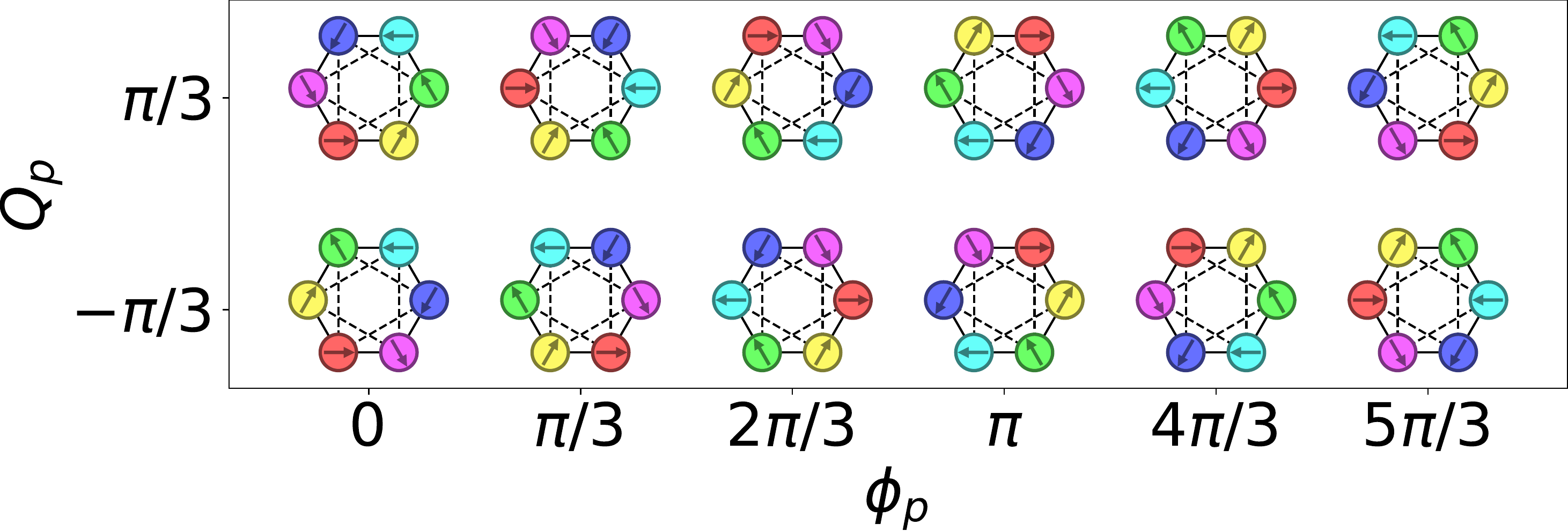}
\end{center}
\caption{Spiral colored states with $Q_p=\pm \pi/3$ and $\phi_p=2\pi m/L_p$ for $m=0,\ldots,L_p-1$ on a hexagonal motif. All of the depicted states are ground states of the same motif Hamiltonian $\hat{H}_p[Q_p]=\hat{H}_p[-Q_p]$ in Eq.~(\ref{eq:HQ}).} \label{fig:hexagons}
\end{figure}

\section{Hamiltonians made from motifs}

Here we describe how to construct Hamiltonians on structures built from motifs with projected spiral colored eigenstates. These Hamiltonians can be frustration-free, in which case the projected spiral colored states are ground states on each motif, or frustrated, in which case the states are excited states.

\begin{figure}
\begin{center}
\includegraphics[width=0.5\textwidth]{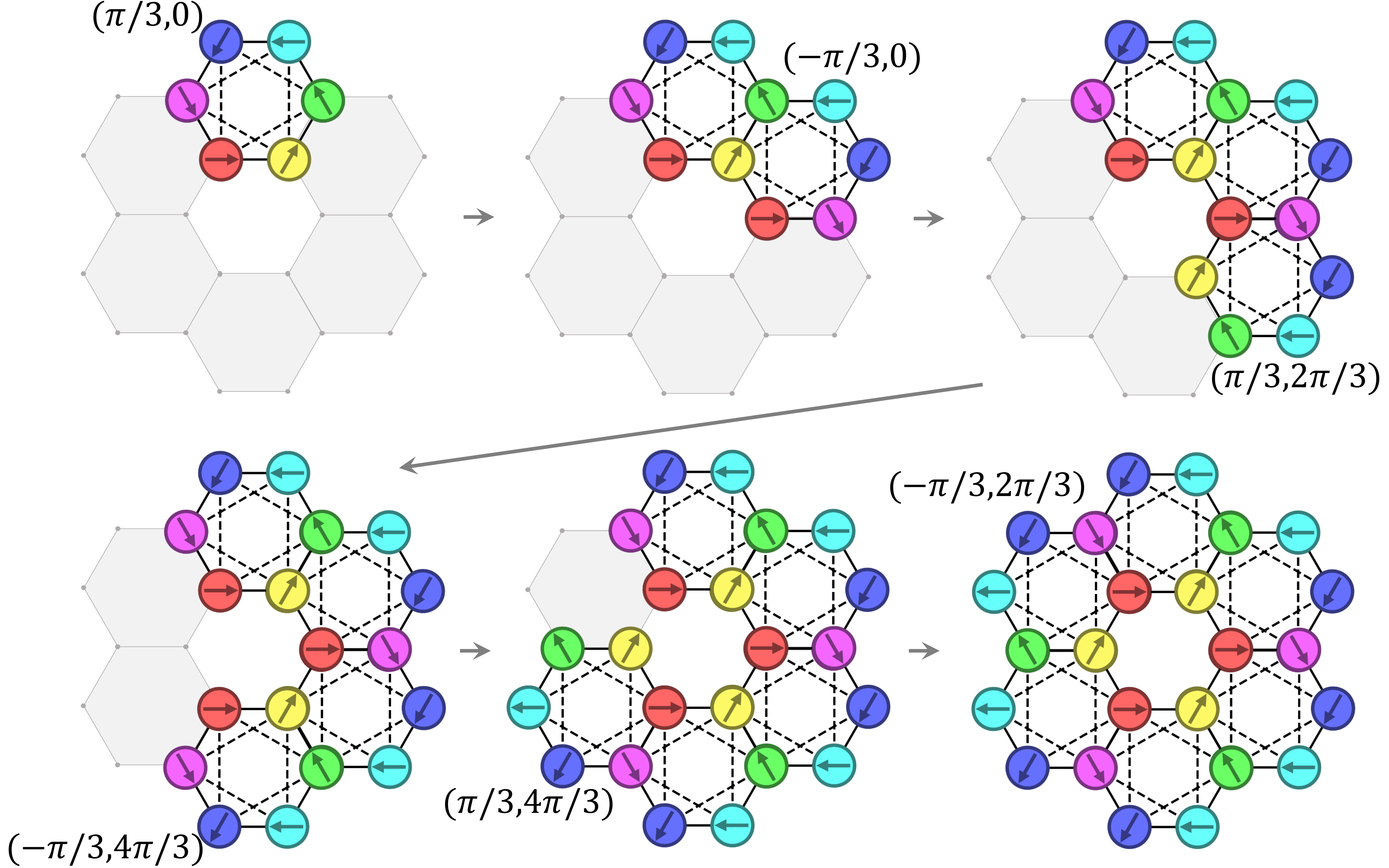}
\end{center}
\caption{An example of how to construct a 6-colored Hamiltonian (Eq.~(\ref{eq:H_Zeeman})) with a spiral colored eigenstate (Eq.~(\ref{eq:psi_coloring})) using a hexagonal motif. Each newly added hexagon motif $p$ is labelled by the $(Q_p,\phi_p)$ for the motif spiral colored state $\ket{\psi_p[Q_p,\phi_p]}$ (Eq.~(\ref{eq:psiQ})) on that hexagon (See also Fig.~\ref{fig:hexagons}). Note that no motif Hamiltonian is added to the center hexagon since that hexagon cannot be 6-colored in the appropriate way.} \label{fig:motifs_to_lattice}
\end{figure}

Consider a geometric structure composed of motifs, such as a triangular lattice tiled by triangles. By taking a linear combination of Hamiltonians on those motifs, we can build a Hamiltonian on the entire system. We focus on Hamiltonians of this form with an additional Zeeman term
\begin{align}
\hat{H} = \sum_p J_p \hat{H}_p[Q_p] - h \sum_i \hat{\sigma}^z_i. \label{eq:H_Zeeman}
\end{align}
In order for Eq.~(\ref{eq:H_Zeeman}) to have projected spiral colored eigenstates, the $\hat{H}_p[Q_p]$ need to be added together in such a way so that the spiral colored product states $\ket{\psi_p[Q_p,\phi_p]}$ (Eq.~(\ref{eq:psiQ})) on overlapping motifs $p$ agree. In practice, the rules for adding together motif Hamiltonians can be understood through visual inspection: spiral-colored motifs can only be added together if the colors on spins shared between neighboring motifs match (see Fig.~\ref{fig:motifs_to_lattice}).

First, suppose that $h=0$. Then the spiral colored states
\begin{align}
\ket{C} = \bigotimes_{j} \ket{\gamma_j}_j, \label{eq:psi_coloring}
\end{align}
where $\ket{\gamma_j}_j$ in state $\ket{\psi_p[Q_p,\phi_p]}$ must be the same on each motif $p$ containing site $j$, are eigenstates of Eq.~(\ref{eq:H_Zeeman}). There can be many $\ket{C}$ because of the degeneracy between $\ket{\psi_p[Q_p,\phi_p]}$ and $\ket{\psi_p[-Q_p,\phi_p']}$ on each motif. For example, in Ref.~\onlinecite{Changlani2017}, the authors found that certain lattices built from triangle motifs, such as the kagome lattice, have exponentially many different 3-colorings, i.e., exponentially many linearly independent $\ket{C}$. When $J_p \geq 0$ for all motifs $p$, then the spiral colored states $\ket{C}$ are ground states of $J_p \hat{H}_p$ and $\hat{H}$, making Eq.~(\ref{eq:H_Zeeman}) a frustration-free Hamiltonian. If, however, the $J_p$ alternate in sign, then $\ket{C}$ are excited states of $J_p \hat{H}_p$ and $\hat{H}$ and Eq.~(\ref{eq:H_Zeeman}) is frustrated.

Next, suppose that $h > 0$. This breaks the degeneracies in different $S_z$ sectors and makes the product states $\ket{C}$ no longer eigenstates. Instead, the \emph{projected} product states
\begin{align}
\ket{C_{S_z}} \equiv \frac{\hat{P}_{S_z} \ket{C}}{\norm{\hat{P}_{S_z} \ket{C}}} = \frac{1}{\mathcal{N}_{S_z}}\hat{P}_{S_z} \ket{C}, \label{eq:C_Sz}
\end{align}
where $\mathcal{N}_{S_z}$ is a $S_z$-sector normalization constant, are eigenstates of Eq.~(\ref{eq:H_Zeeman}) with energies $\sum_p J_p E_p - h S_z$. Note, however, that many of the projected spiral colored states are \emph{not} linearly independent. For example, $\hat{P}_{S_z=N}\ket{C}=\rket{\uparrow\cdots\uparrow}$ for all $\ket{C}$ where $N$ is the number of spins in the system. If the Hamiltonian is frustration-free so that $J_p \geq 0$ for all motifs $p$, then $\rket{\uparrow\cdots\uparrow}$ is the ground state and the $\ket{C_{S_z}}$ are the lowest energy states in each $S_z$ sector. If the Hamiltonian is frustrated so that the $J_p$ alternate in sign, then the $\ket{C_{S_z}}$ are excited states. Ref.~\onlinecite{Lee2020} showed that projected product states of this form have $\sim \log N$ entanglement entropy, which indicates that the $\ket{C_{S_z}}$ are athermal and for frustrated Hamiltonians in Eq.~(\ref{eq:H_Zeeman}) can be quantum many-body scars.

\section{Examples of spiral colored systems}

The Hamiltonian in Eq.~(\ref{eq:H_Zeeman}) and its associated spiral colored eigenstates can be realized in many different lattice geometries built out of motifs. In this section we describe how this can be done. First, we begin by analyzing in more detail the properties of the $n$-spin motifs listed in Table~\ref{tab:HQ}. Then, we provide examples of different geometries that can be constructed from these motifs.

\emph{Bonds.} The simplest motif is the 2-spin motif, or bond. In this case, the 2-spin motif Hamiltonian is simply $\hat{H}_p[Q_p]=\mp(\hat{\sigma}_1^x \hat{\sigma}_2^x + \hat{\sigma}_1^y \hat{\sigma}_2^y) - \hat{\sigma}_1^z \hat{\sigma}_2^z$ for $Q_p=0,\pi$. 

When $Q_p=0$, this is a ferromagnetic Heisenberg bond and the (1-colored) spiral colored state $(\rket{\uparrow}_1 + e^{i\phi_p}\rket{\downarrow}_1)(\rket{\uparrow}_2 + e^{i\phi_p}\rket{\downarrow}_2)$, which corresponds to two aligned spins in the $xy$-plane, is a ground state. Note that this implies that $\rket{+}_i\otimes\rket{+}_j$ is an eigenstate of any Heisenberg bond $\mathbf{S}_i\cdot \mathbf{S}_j$ with energy\footnote{The Heisenberg bond has energy eigenvalues $-3/4,1/4,1/4,1/4$, so $\rket{++}$ is in the degenerate $+1/4$ eigenspace of the operator.} $+1/4$. This also implies that $\rket{+\cdots +}$ is an exact eigenstate of any Hamiltonian made of only Heisenberg bonds. If all the Heisenberg bonds are ferromagnetic (antiferromagnetic), then $\ket{+\cdots +}$ is the ground state (highest excited state). However, if some bonds are ferromagnetic and some are antiferromagnetic, then $\rket{+ \cdots +}$ (and by symmetry $\rket{\chi \cdots \chi}$ for any single-spin state $\rket{\chi}$) would be an eigenstate in the middle of the spectrum and would likely be a quantum many-body scar state.

When $Q_p=\pi$, the 2-spin motif Hamiltonian is an antiferromagnetic XXZ$[-1]$ bond and the (2-colored) state $(\rket{\uparrow}_1 + e^{i\phi_p}\rket{\downarrow}_1)(\rket{\uparrow}_2 + e^{i(\phi_p+\pi)}\rket{\downarrow}_2)$, which corresponds to two anti-aligned spins in the $xy$-plane, is a ground state. The 2-colored XXZ$[-1]$ model is also discussed in Ref.~\onlinecite{Pal2021}.

\begin{figure}
\begin{center}
\includegraphics[width=0.45\textwidth]{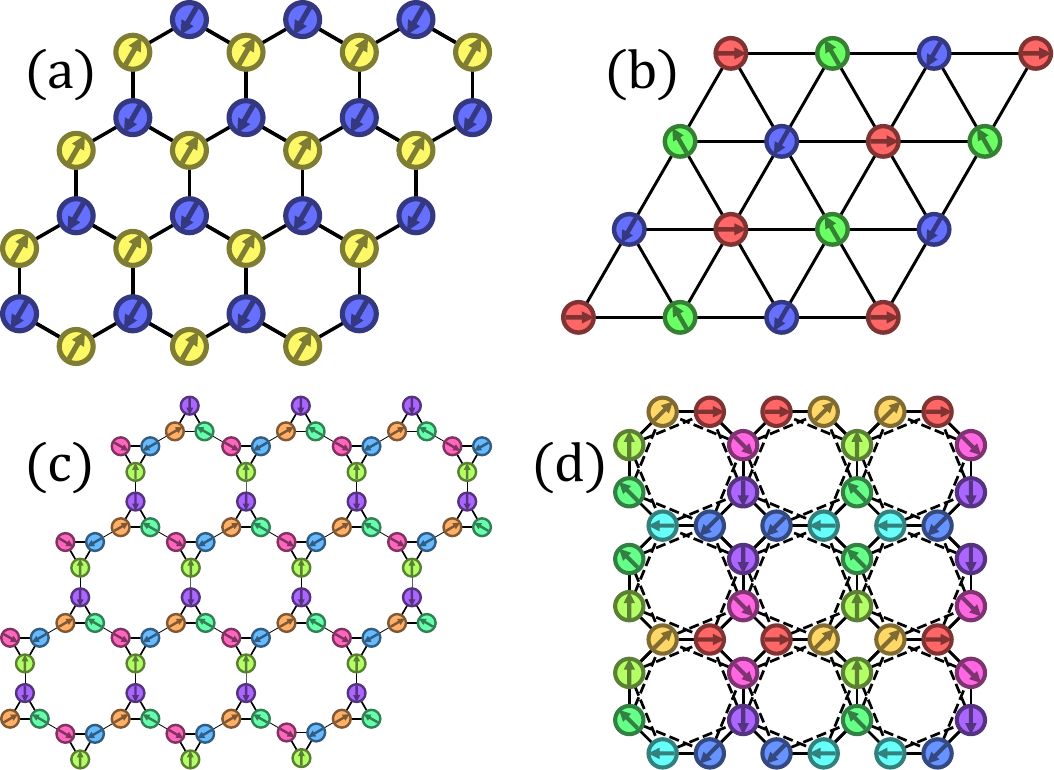}
\end{center}
\caption{Spiral colored Archimedean tilings. (a) A 2-colored honeycomb lattice made from bond motifs. (b) A 3-colored triangular lattice made from triangle motifs. (c) A 2- and 3-colored truncated hexagonal tiling lattice made from bond and triangle motifs. (d) An 8-colored square-octagon lattice made from octagon motifs.} \label{fig:archimedean_lattices}
\end{figure}

The $Q_p=0$ and $Q_p=\pi$ bonds are quantum analogs of ferromagnetic and antiferromagnetic Ising bonds and promote order analogous to ferromagnetic and N\'{e}el order, though in the $xy$-plane. When ``spiral coloring'' a lattice, the $Q_p=0$ bond can be added between any two aligned spins and the $Q_p=\pi$ bond can be added between any two anti-aligned spins. By itself, the $Q_p=0$ bond can 1-color any lattice. Similarly, by itself, the $Q_p=\pi$ bond can 2-color any bipartite graph, such as square, honeycomb lattices (see Fig.~\ref{fig:archimedean_lattices}(a)), and pyrochlore lattices.

\emph{Regular polygons.} The next simplest motifs are regular polygons, shown in Fig.~\ref{fig:polygons}. The smallest example is the equilateral triangle, made of $L_p=3$ sites. When $Q_p=\pm 2\pi/3$, the triangle motif Hamiltonian is made of three $XXZ[-1/2]$-bonds and is known as the XXZ0 model \cite{Essafi2016,Changlani2017}. It has 3-colored ground states $(\rket{\uparrow}_1 + e^{i\phi_p}\rket{\downarrow}_1)(\rket{\uparrow}_2 + e^{i(\phi_p\pm 2\pi/3)}\rket{\downarrow}_2)(\rket{\uparrow}_3 + e^{i(\phi_p\mp 2\pi/3)}\rket{\downarrow}_3)$, that correspond to spins at $120^{\circ}$ angles from one another in the $xy$-plane. The 3-colored state has a unique property among the spiral colored states on polygons: all permutations of the three colors ($\ket{012},\ket{021},\ket{102},\ldots$) produce valid spiral colored states. This property does not hold for example for a spiral colored state on a hexagon, where for instance $\ket{021345}$ would not be a valid spiral colored state. Note also that there is a constraint that occurs when building lattices out of edge-sharing polygon motifs: $Q_p=-Q_{p'}$ if polygons $p$ and $p'$ share an edge. In some lattices, this constraint cannot be satisfied unless some motifs in the lattice are missing, as can be seen in Fig.~\ref{fig:motifs_to_lattice}. Finally, note that 4-colored square motifs have vanishing nearest neighbor interactions (see~Table~\ref{tab:HQ}), so cannot be used to construct Hamiltonians with 4-colored spiral colored eigenstates.

\begin{figure}
\begin{center}
\includegraphics[width=0.45\textwidth]{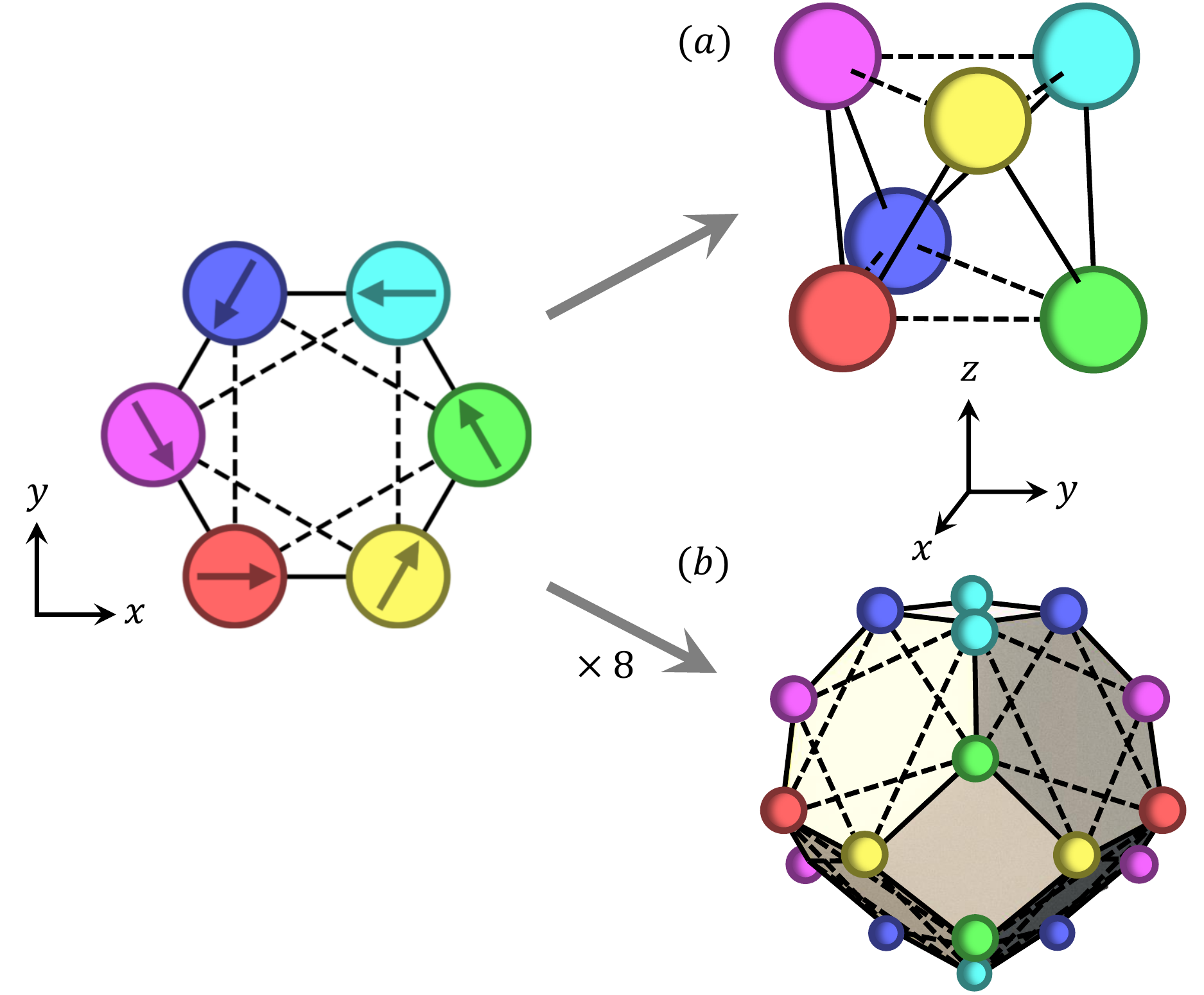}
\end{center}
\caption{Spiral colored states and Hamiltonians on a three-dimensional (a) anti-prism (octahedron) motif and (b) a truncated octahedron motif made from a two-dimensional polygon (hexagon) motif.} \label{fig:polyhedra}
\end{figure}

\emph{Polyhedra.} Motifs involving four or more spins can be reshaped in three-dimensions into polyhedra. A natural type of polyhedron motif to consider is an antiprism, a polyhedron that is composed of two parallel copies of a regular polygon connected by an alternating band of triangles. The nearest and next-nearest neighbor interactions in the motif Hamiltonian of Eq.~(\ref{eq:HQ}) act between and within the two parallel polygons, respectively. Fig.~\ref{fig:polyhedra}(a) shows an example of an antiprism motif (an octahedron motif). Other types of polyhedra motifs are those that can be built out of polygon motifs. For example, a truncated octahedron motif can be made out of eight hexagon motifs, as shown in Fig.~\ref{fig:polyhedra}(b).

\begin{figure}
\begin{center}
\includegraphics[width=0.45\textwidth]{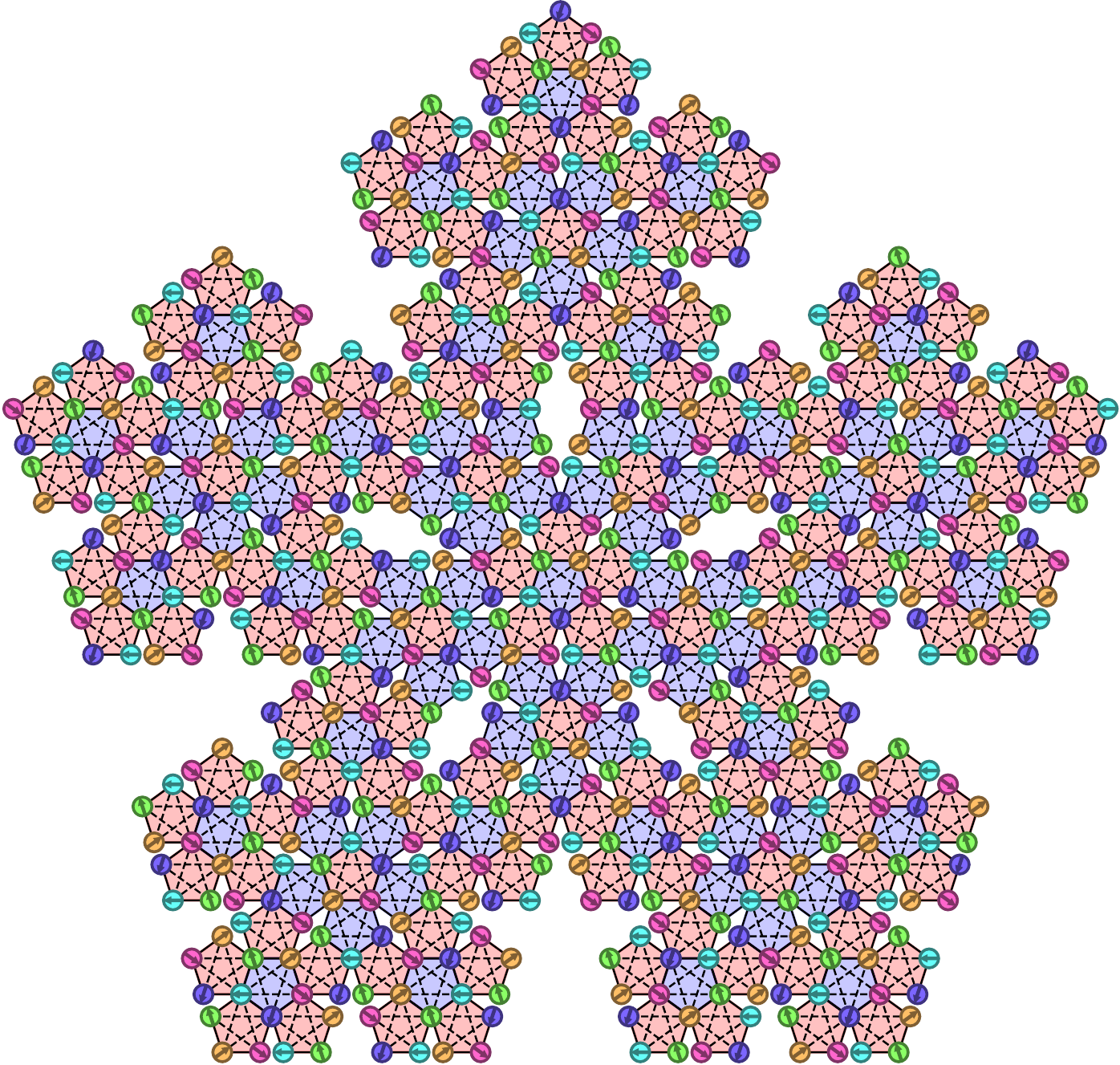}
\end{center}
\caption{A spiral colored state on an aperiodic Penrose diagram made of pentagon motifs. The shaded pentagons have nearest neighbor ferromagnetic $XXZ[1/2\varphi]$ and next-nearest neighbor antiferromagnetic $XXZ[-\varphi/2]$ interactions, where $\varphi=(1+\sqrt{5})/2$ is the golden ratio. Purple (red) pentagons have $Q_p=+2\pi/5$ ($Q_p=-2\pi/5$).} \label{fig:penrose_tiling}
\end{figure}

Polygon and polyhedron motifs can be combined into many different geometries. Here we list some examples of different geometries of spin systems that can be spiral colored using the motifs that we have described above:

\begin{enumerate}
    \item \emph{Archimedean tilings.}  Archimedean tilings are tilings of the Euclidean plane by regular polygons \cite{Conway2008}. Examples of spiral colored Archimedean tilings are shown in Fig.~\ref{fig:archimedean_lattices}.
    \item \emph{Honeycombs.} Space-filling tesselations of three-dimensional space, or honeycombs, can be constructed out of certain polyhedra \cite{Conway2008}. For example, a bitruncated cubic honeycomb can be made up of truncated octahedra.
    \item \emph{Lattices with polygons and polyhedra in the unit cell.} The lattices of crystalline materials can have unit cells with polygons or polyhedra in them. For example, in frustrated magnets, the spins can be arranged into a two-dimensional kagome lattice, which has two triangles in a unit cell, or a three-dimensional pyrochlore lattice, which has tetrahedra in its unit cell.
    \item \emph{Aperiodic tilings.} Aperiodic patterns, such as those seen in quasicrystals \cite{Schechtman1984,Levine1984}, can be built out of simple motifs. Fig.~\ref{fig:penrose_tiling} shows an example of an aperiodic spiral colored Penrose tiling \cite{Penrose1974} made from pentagon motifs.
\end{enumerate}

\section{Numerical signatures of quantum many-body scars}

In this section, we provide numerical evidence that the projected spiral colored states $\ket{C_{S_z}}$ are quantum many-body scars of frustrated Hamiltonians in Eq.~(\ref{eq:H_Zeeman}) and that the scars give rise to periodic revivals when time-evolving from the unprojected $\ket{C}$ state. (See Ref.~\onlinecite{Lee2020} for a similar analysis for 3-colored states.) We use exact diagonalization (ED) and the time-evolved block decimation (TEBD) algorithm \cite{Vidal2003} to analyze the eigenstates and study the dynamics of these scar Hamiltonians.

\begin{figure}
\begin{center}
\includegraphics[width=0.45\textwidth]{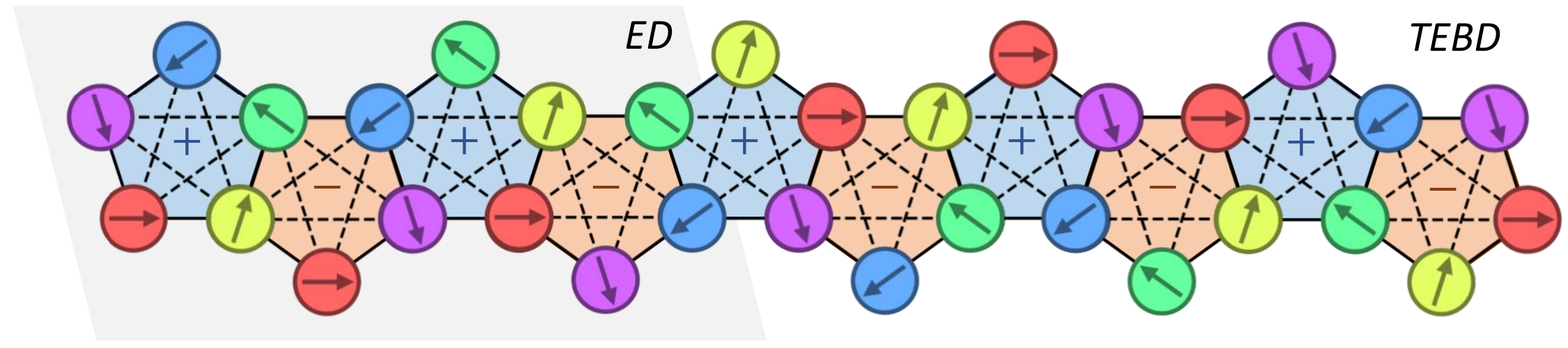}
\end{center}
\caption{A spiral colored state $\ket{C}$ on a chain of pentagons. Projected spiral colored states $\ket{C_{S_z}}\propto \hat{P}_{S_z}\ket{C}$ are quantum many-body scars of the Hamiltonian in Eq.~(\ref{eq:H_scar}). Positive (negative) terms in the Hamiltonian are indicated by blue and a ``$+$'' (red and a ``$-$''). Our exact diagonalization (ED) calculations were done on the 14 spins in the first 4 pentagons, highlighted in gray. Our time-evolved block decimation (TEBD) calculations were done on the 32 spins in the 10 pentagons shown.} \label{fig:pentagon_chain}
\end{figure}

Consider the pentagon chain Hamiltonian in a magnetic field
\begin{align}
\hat{H}=\sum_{j=1}^{M}(-1)^{j-1}\hat{H}_{p_j}[2\pi/5] - h\sum_{i=1}^N \hat{\sigma}^z_i \label{eq:H_scar}
\end{align}
where $p_1,\ldots,p_M$ are edge-sharing pentagons arranged into a line as shown in Fig.~\ref{fig:pentagon_chain}. This chain has open boundary conditions and $N=3M+2$ spins. The motif Hamiltonians $\hat{H}_{p_j}[2\pi/5]$ are normalized according to the $L_p=5$ row in Table~\ref{tab:HQ}. Note that the nearest neighbor bonds shared between pentagons exactly cancel. Also, note that $\hat{H}$ anticommutes with the $180^\circ$ spatial rotation operator $\hat{R}$, which can be seen by rotating the pentagon chain about its center and noting that the ``$+$'' and ``$-$'' signs swap so that $\hat{H}\rightarrow -\hat{H}$. Since $\{ \hat{H}, \hat{R}\}=0$, each energy $E$ eigenstate $\ket{E}$ has an associated energy $-E$ eigenstate $\hat{R}\ket{E}$.

In our analysis of this model, we will focus on a particular spiral colored state $\ket{C}$ shown in Fig.~\ref{fig:pentagon_chain}. Because of the magnetic field, $\ket{C}$ is not an energy eigenstate of the Hamiltonian, but the projected $\ket{C_{S_z}}$ states are, with energies $-hS_z$. The $\ket{C_{S_z}}$ states are in the middle of the spectrum of their own $S_z$ sectors, suggesting that they are scar states. Since $\ket{C}$ is a linear combination of these states, it inherits their athermal properties. To verify the nature of the scar states in this system, we will check that: (1) generic energy eigenstates of Eq.~(\ref{eq:H_scar}) are thermal, (2) the $\ket{C_{S_z}}$ states are athermal, and (3) that the time evolution of the product state $\ket{C}$, which only has non-zero weight on scar states, exhibits periodic revivals unlike the time evolution of a random product state, which rapidly thermalizes.

\begin{figure}
\begin{center}
\begin{tabular}{c}
\includegraphics[width=0.37\textwidth]{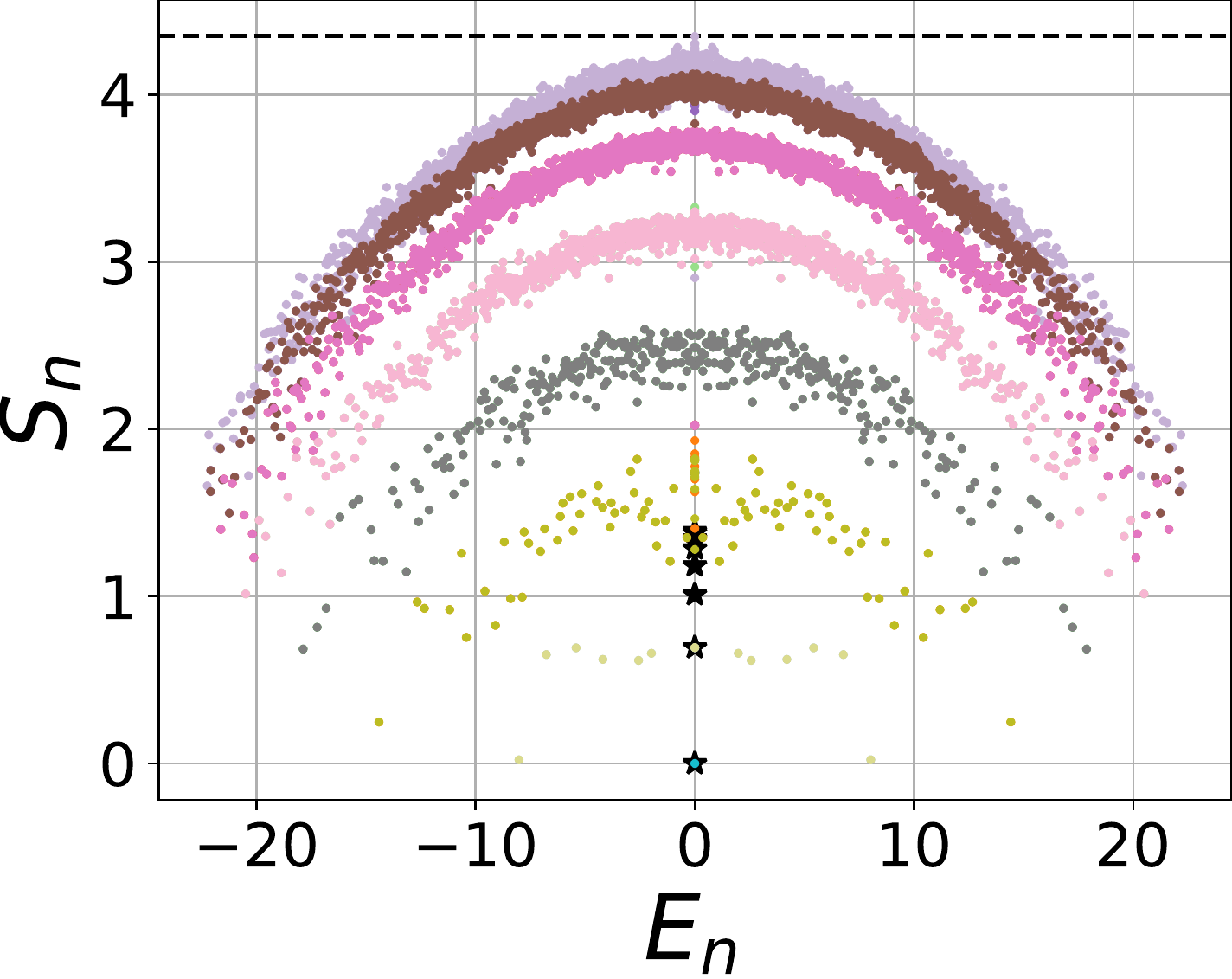} \\
\includegraphics[width=0.37\textwidth]{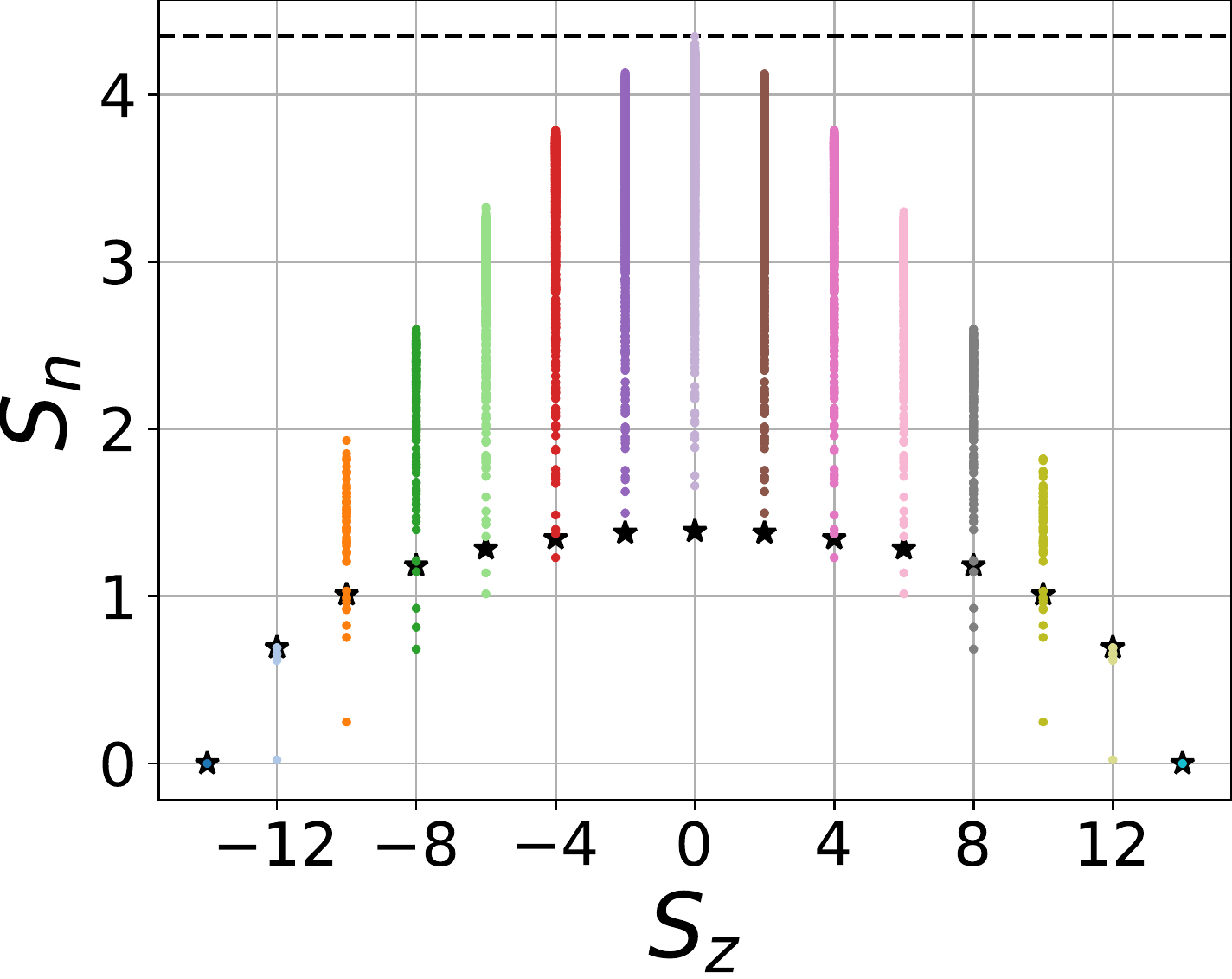}
\end{tabular}
\end{center}
\caption{\textbf{Top:} The half-cut entanglement entropy $S_n$ of the energy eigenstates of $\ket{\psi_n}$ of the pentagon chain Hamiltonian in Eq.~(\ref{eq:H_scar}) with $h=0$ versus energy $E_n$, computed using ED on 14 spins. Different colors correspond to different $S_z$ sectors. \textbf{Bottom:} Entropy $S_n$ versus $S_z$ quantum number. In both plots, the black stars correspond to the entropies of the projected spiral colored states $\ket{C_{S_z}}$ and the black dashed line corresponds to the average half-cut entropy of a Haar-random pure state.} \label{fig:pentagon_chain_entropies}
\end{figure}

\begin{figure}
\begin{center}
\includegraphics[width=0.37\textwidth]{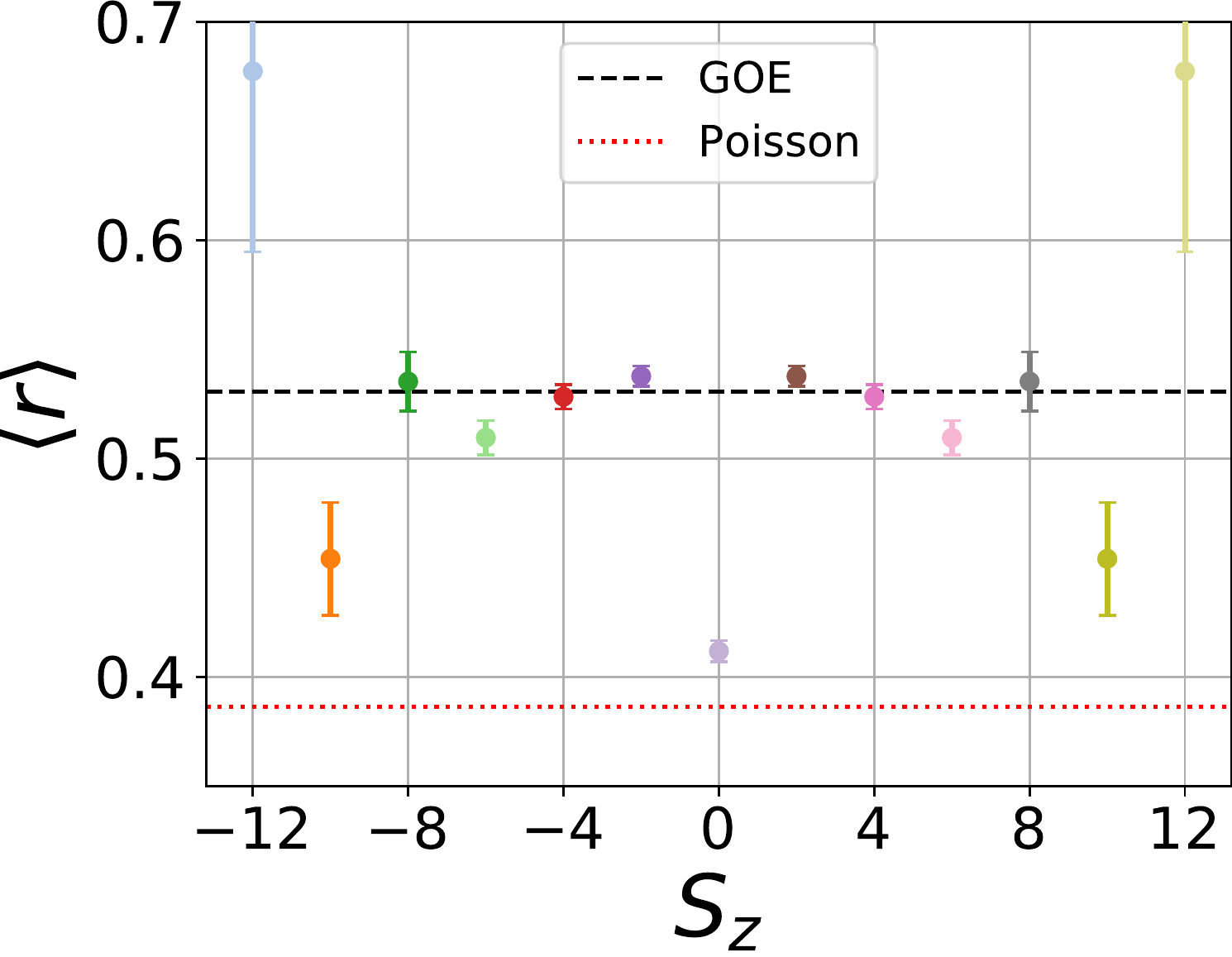} 
\end{center}
\caption{The average level-spacing ratio $\langle r \rangle$ in each $S_z$ quantum number sector of the pentagon chain Hamiltonian depicted in Fig.~\ref{fig:pentagon_chain}, computed using ED on 14 spins. The error bars are standard errors of the mean.} \label{fig:pentagon_chain_level_spacing_ratios}
\end{figure}

First, we use full ED on the 14-spin pentagon chain shown in Fig.~\ref{fig:pentagon_chain} to compute the properties of the energy eigenstates. Fig.~\ref{fig:pentagon_chain_entropies} shows the von-Neumann half-chain entanglement entropy $S_n = -\textrm{tr}(\rho_n \log \rho_n)$ for each eigenstate $\ket{\psi_n}$ [$\rho_n = \textrm{tr}_{\{j=N/2+1,\ldots,N\}}(\ket{\psi_n}\bra{\psi_n})$] in each $S_z$ sector. The half-cut entropies of the $\ket{C_{S_z}}$ states are marked with black stars and the average half-cut Page entropy for a Haar-random pure state\cite{Page1993} $S_{avg}=N/2 \log 2 - 1/2$ is marked with a black dashed line for reference. The entropies of the $\ket{\psi_n}$ eigenstates with $E$ and $S_z$ near zero are close to the Page entropy. We also compute the average level-spacing ratio $r_n = \frac{\min (E_n - E_{n-1}, E_{n+1} - E_n)}{\max (E_n - E_{n-1}, E_{n+1} - E_n)}$, where $E_n$ are the ordered energy eigenvalues, for the energy eigenstates in each $S_z$ sector. For an ergodic Hamiltonian, $\langle r \rangle$ is expected to be near the Gaussian orthogonal ensemble (GOE) value of $r_{GOE}=0.5307$, while for an integrable model, it is expected to be near the Poisson value of $r_{Poisson} = 0.3863$. We find that $\langle r \rangle_{S_z}$ are within a few error bars of $r_{GOE}$ for $S_z=\pm 8,6,4,2$, but not $S_z=0$ (see Fig.~\ref{fig:pentagon_chain_level_spacing_ratios}). We believe that this is due to a hidden symmetry in the $S_z=0$ sector that we did not take into account in our analysis, which can make $\langle r \rangle$ approach the $r_{Poisson}$ value. The large entanglement of generic finite energy density eigenstates and the GOE level-spacing statistics in most $S_z$ sectors provide strong evidence that Eq.~(\ref{eq:H_scar}) is an ergodic Hamiltonian with mostly thermal eigenstates.

Next, we compute the half-cut entanglement entropies of the energy eigenstates and the $\ket{C_{S_z}}$ states, shown in Fig.~\ref{fig:pentagon_chain_entropies}. Through ED, we find that the 14-spin Hamiltonian has a degeneracy of $1,2,9,2,23,2,37,128,37,2,23,2,9,2,1$ at zero energy for $S_z=-14,\ldots,14$. Due to this degeneracy, the zero energy eigenstates obtained through numerical diagonalization can be arbitrary linear combinations of the degenerate states and have ill-defined entropies, which is why there is not perfect agreement between the numerical and exact $\ket{C_{S_z}}$ eigenstates in Fig.~\ref{fig:pentagon_chain_entropies}. Nonetheless, we numerically verify that the $\ket{C_{S_z}}$ states are exact eigenstates of Eq.~(\ref{eq:H_scar}) with energies in the middle of the spectra and with low entanglement entropy far from the Page value. This indicates that the $\ket{C_{S_z}}$ are athermal eigenstates.

\begin{figure}
\begin{center}
\includegraphics[width=0.4\textwidth]{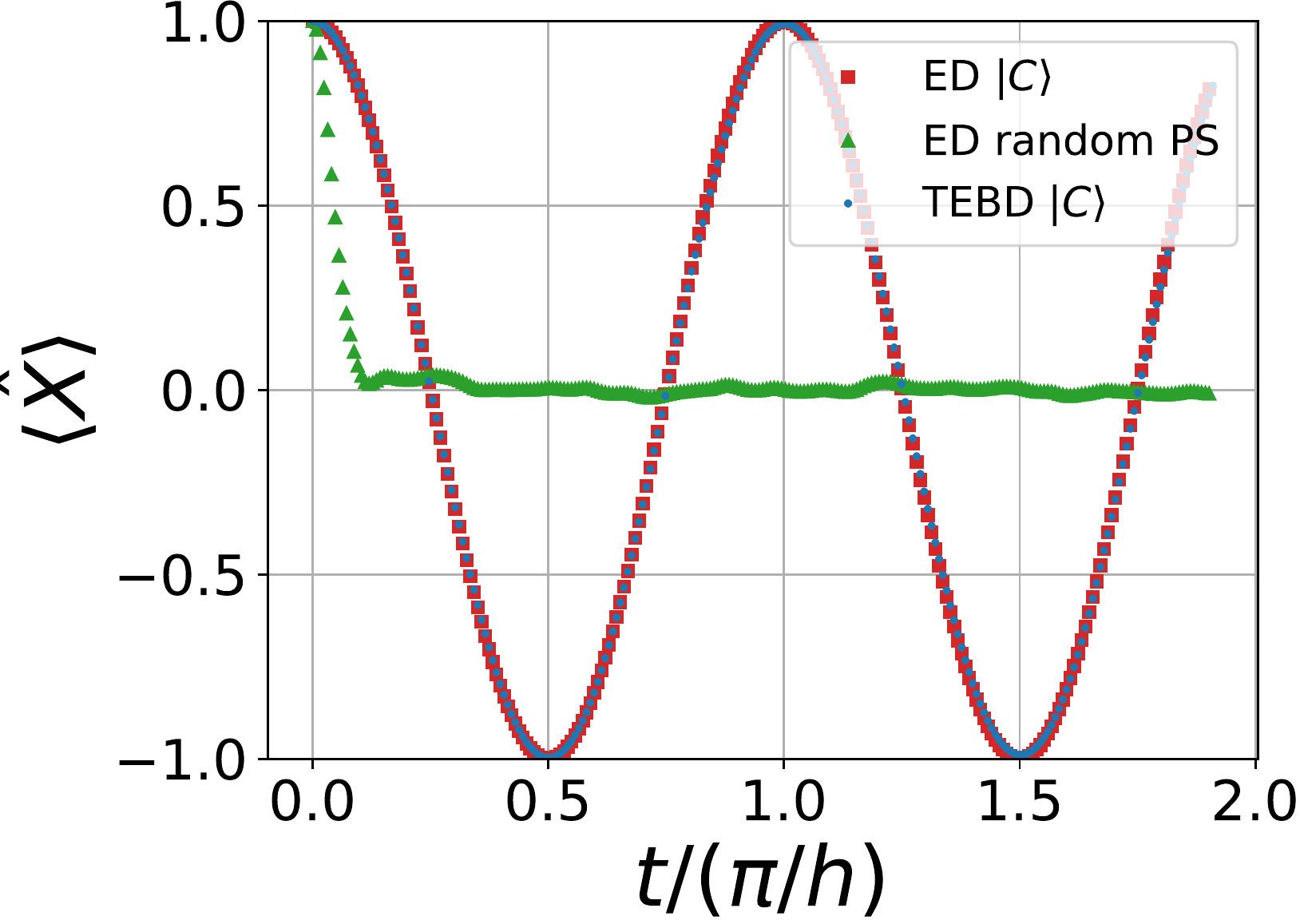}
\end{center}
\caption{The expectation value of $\hat{X} = \hat{\sigma}^x_6$ under the pentagon chain Hamiltonian with $h=0.5$ using $\ket{C}$ and a random product state whose 6th spin is set to $\ket{+}$ as initial conditions. The $\ket{C}$ evolution was computed using ED for 14 spins and TEBD for 32 spins.} \label{fig:pentagon_chain_time_evolution}
\end{figure}

Finally, we examine the time-evolution of our system with different initial product states to see how the athermal nature of the scar states manifests itself in dynamics that can be observed experimentally. First, we consider the initial state $\ket{C}$. Since it is a superposition of scar states $\ket{C_{S_z}}$ at different $S_z$ with energies $-hS_z$, the time-evolved state will be
\begin{align}
\ket{\psi(t)}=e^{-it\hat{H}}\ket{\psi(0)} = \sum_{S_z} \mathcal{N}_{S_z} e^{ithS_z}\ket{C_{S_z}}.
\end{align}
This state exhibits periodic revivals, which can be detected by measuring local observables. For example, the $x$-magnetization of a single spin will oscillate
\begin{align}
\bra{\psi(t)}\hat{\sigma}_i^x \ket{\psi(t)}= \sum_{S_z,S_z'} \mathcal{N}_{S_z'}^*\mathcal{N}_{S_z} e^{ith(S_z-S_z')} \bra{C_{S_z'}}\hat{\sigma}_i^x \ket{C_{S_z}} \label{eq:X_mag}
\end{align}
with a period $T=\frac{2\pi}{h \Delta S_z} = \frac{\pi}{h}$ since $\bra{C_{S_z'}}\hat{\sigma}_i^x \ket{C_{S_z}} \neq 0$ only when $S_z'=S_z\pm 2$. As discussed in the appendix, during the time-evolution, the spins in $\ket{C}$ precess around the $z$-axis at frequency $2\pi/T$ so $\ket{C(t)}=e^{-it\hat{H}}\ket{C}$ remains as a product state.

Using ED on the 14-spin pentagon chain and TEBD on the 32-spin chain, we compute Eq.~(\ref{eq:X_mag}) for the $i=6$ spin, which is initially in the $\ket{+}$ state (see Fig.~\ref{fig:pentagon_chain_time_evolution}). We clearly see the characteristic revival with period $T=\pi/h$. Next, we consider using a random product state whose $i=6$ spin is set to $\ket{+}$ as the initial state. As shown in Fig.~\ref{fig:pentagon_chain_time_evolution}, the same observable for the random product state quickly decays to zero, signaling thermalization. 

\section{Conclusions and outlook}

We introduced a large family of Hamiltonians with exact projected spiral colored eigenstates built out of $n$-spin motifs. We showed that these Hamiltonians can be frustration-free so that these states are ground states or frustrated so that they are quantum many-body scars. 

The spiral colored Hamiltonians have simple two-local interactions, can be realized in many different geometries, and generate periodic revival dynamics for simple initial product states. These features make these models promising candidates for experimental observation of quantum many-body scars. One option would be to find a magnetic material with spiral colored physics and observe its dynamics from an initially polarized state in an applied magnetic field \cite{Lee2020}. Another option would be to simulate the periodic revival dynamics of these spiral colored Hamiltonians using a quantum simulator \cite{Georgescu2014} or a digital quantum computer \cite{Preskill2018}. In these dynamics experiments, one could also examine how robust the periodic revivals are to different perturbations to the Hamiltonian.

It would also be interesting to explore the connections of these spiral colored Hamiltonians with frustrated magnetism and spin liquid physics. It is possible that previously studied two and three-dimensional magnetic Hamiltonians are close enough to the spiral colored models so that they inherit some of their low-temperature properties from the spiral colored states. It is also possible that there exist models on lattices with exponentially many spiral colored ground states, like the 3-colored kagome model with exponentially degenerate 3-colorings, which could potentially be close to a quantum spin liquid phase.

Finally, the techniques developed here could be used to find new families of Hamiltonians with exact scar states. For example, by using motif Hamiltonians with valence bond ordered ground states, such as the Majumdar-Ghosh model \cite{Majumdar1969,Majumdar1970}, families of Hamiltonians with valence bond solid scar states could be built. More generally, motif Hamiltonians with specific desired eigenstates could be built using eigenstate-to-Hamiltonian algorithms \cite{Qi2019,Chertkov2018,Greiter2018} or parent Hamiltonian techniques \cite{PerezGarciaI2007,PerezGarciaII2007} and could be used to design many new families of scar models built from motifs.

\begin{acknowledgments}

\emph{Acknowledgments.}--- We acknowledge useful discussions with Hitesh Changlani. Our TEBD calculations were done using the Julia version of the ITensor library \cite{itensor}. We acknowledge support from the Department of Energy grant DOE de-sc0020165.

\end{acknowledgments}

\appendix

\section{Time-evolution of spiral colored states}

Consider time-evolving the initial product state $\ket{C}$ in Eq.~(\ref{eq:psi_coloring}) by the Hamiltonian $\hat{H}$ in Eq.~(\ref{eq:H_Zeeman}):
\begin{align*}
\ket{C(t)} = e^{-it\hat{H}}\ket{C}
\end{align*}
By construction, $\ket{C}$ is an eigenstate of each motif Hamiltonian $\hat{H}_p[Q_p]$ with energy $E_p$. Also, $[\hat{H}_p, \sum_j \hat{\sigma}^z_j]=0$. Together, these imply that
\begin{align*}
\ket{C(t)} &= e^{ith \sum_j \hat{\sigma}^z_j} e^{-it\sum_p J_p \hat{H}_p} \ket{C} \\
&= e^{-it\sum_p J_p E_p} e^{ith \sum_j \hat{\sigma}^z_j} \ket{C} \\
&= e^{-it\sum_p J_p E_p} \bigotimes_j e^{ith \hat{\sigma}^z_j} \ket{\gamma_j}_j.
\end{align*}
This indicates that, up to a global phase, the time evolution applied to this state causes each spin to precess around the $z$-axis at the same rate.

Consider a particular spin $j$ in $\ket{C}$ initially in state $\ket{\gamma_j}=\rket{+}=(1,1)^T/\sqrt{2}$. After time $t$, this spin's state will be
\begin{align*}
e^{ith \hat{\sigma}^z}\rket{+} = \begin{pmatrix} e^{ith} & 0 \\ 0 & e^{-ith} \end{pmatrix} \frac{1}{\sqrt{2}} \begin{pmatrix} 1 \\ 1 \end{pmatrix} = \frac{e^{ith}}{\sqrt{2}} \begin{pmatrix} 1 \\ e^{-2ith} \end{pmatrix},
\end{align*}
which is $\propto \rket{+}$ when $t=T=\pi/h$.

\bibliography{refs}

\end{document}